\documentclass[reprint,amssymb,amsmath,aip,cha]{revtex4-1}
\usepackage{graphicx}
\usepackage[caption = false]{subfig}
\usepackage{color}
\usepackage{epsfig}
\usepackage{ifpdf}
\usepackage{bm}
\usepackage[colorlinks=true,linkcolor=blue]{hyperref}%
\expandafter\ifx\csname package@font\endcsname\relax\else
 \expandafter\expandafter
 \expandafter\usepackage
 \expandafter\expandafter
 \expandafter{\csname package@font\endcsname}%
\fi
\usepackage{cleveref}
\begin{document}
\title{Response of the low pressure hot-filament discharge plasma to a positively biased auxiliary disk electrode}
\author{Mangilal Choudhary}
\email{jaiijichoudhary@gmail.com} 
\affiliation{Institute of Advanced Research, Koba, Gandhinagar, 382426, Gujarat, India}
\author{P. K. Sreejith}
\affiliation{Indian Institute of Technology Madras, Chennai, Tamil Nadu 600036, India}
%
\begin{abstract}
In steady-state, the plasma loss rate to the chamber wall is balanced by the ionization rate in the hot-filament discharges. This balance in the loss rate and ionization rate maintains the quasi--neutrality of the bulk plasma. In this report, we have studied the properties of bulk plasma in the presence of an auxiliary (additional) biased metal disk, which is working as a sink, in low-pressure helium plasma. A single Langmuir probe and emissive probe are used to characterize the plasma for various biases (positive and negative) to the metal disk, which was placed along the discharge axis inside the plasma. It is observed that only a positively biased disk increases the plasma potential, electron temperature, and plasma density. Moreover, the plasma parameters remain unaltered when the disk is negatively biased. The observed results for two different-sized positively metal disks are compared with an available theoretical model and found an opposite behavior of plasma density variation with the disk bias voltages at given discharge condition. The role of the primary energetic electron population in determining the plasma parameters is discussed. These experimental results are qualitatively explained on the basis of electrostatic confinement arising due to the loss of electrons to a biased metal disk electrode.
\end{abstract} 
\maketitle
\section{Introduction}
The plasma, which is termed as the fourth state of the matter, is a collection of negatively charged electrons and positively charged ions. These charged species (electrons and ions) are not free but strongly affected by the electromagnetic field of surrounding charged species and capable of exhibiting the collective response. In laboratory plasma, a fraction of neutrals is present along with the charged species and the ionization fraction depends on the density of neutrals. However, the plasma is resulting from the ionization of neutral atoms; therefore, it generally contains nearly equal numbers of positive and negative charge carriers and is termed as quasi-neutral plasma. In hot-filament discharges,  the emitted energetic electrons from heated filaments are the source of ionization, and the chamber wall is a sink for the plasma losses. In steady-state, the ionization rate (plasma production rate) and loss rate are same to hold the quasi-neutrality of plasma \cite{chenplasmaphysicsbook,bittencourtplasmaphysicsbook,particlebalancemodel1}.\par
The response of weakly ionized hot-filament discharge to a biased auxiliary metal electrode above the plasma potential gives some very interesting features such as formation of fire ball\cite{fireballstenzel1,fireballstenzel2}, excitation of solitary electron hole \cite{satyasolitaryelectronhole,satyatransientrespons}, electrostatic confinement\cite{electrostaticconfinment1,globalambipolardiffusion}, plasma potential locking\cite{plasmapotentiallocking,anodesizeplasmareponse} etc. At a given discharge condition, the size of an auxiliary metal electrode (diameter in case of disk electrode) and bias voltage above the plasma potential mainly determine the characteristics of plasma. The ratio of area, $A_d/A_w$, determines whether an electron sheath ($A_d/A_w$ $< \mu$) or a double sheath ($A_d/A_w$ $\simeq$ 1.7 $\mu$) or ion sheath ($A_d/A_w$ $>$ 1.7 $\mu$) will be formed around a positively biased ($V_b > V_p$) auxiliary electrode (disk electrode) in plasma volume \cite{plasmapotentiallocking,globalambipolardiffusion,anodesizeplasmareponse}. Here $A_d$ is area of disk electrode, $A_w$ is the plasma facing area of the vacuum chamber, $\mu$ = $\sqrt{2.3 m_e/M_i}$ and $V_b$ is the disk bias voltage. Hence characteristics of bulk plasma strongly depend on the ratio $A_d/A_w$ to $\mu$. The earlier work performed in double plasma device (HOT--filament discharge) suggests lowering the plasma density if the biased voltage of the disk is increased above the plasma potential\cite{globalambipolardiffusion,anodesizeplasmareponse}. \par
In our recent experiment \cite{mangilalcpp}, we observed the plasma density enhancement after application of transient high voltage positive pulses to the axillary disk electrode in low pressure helium plasma. It is expected to observe the plasma density enhancement if a positively biased ($V_b > V_p$) disk electrode is immersed in the hot-filament helium discharge. Furthermore, the experimental device used in our study \cite{mangilalcpp} was different than the double plasma device \cite{anodesizeplasmareponse,plasmapotentiallocking} therefore plasma response to a positively biased auxiliary electrode (disk) should be different than that reported in references\cite{plasmapotentiallocking,globalambipolardiffusion,anodesizeplasmareponse}. Some of these open questions motivate us to perform experiments in the hot-filament helium discharge to see the effect of the positively biased auxiliary disk electrode on the plasma characteristics. \par
In the present investigation, a metallic electrode in the form of disk immersed in unmagnetized helium plasma was biased with a DC voltage source and the response of plasma was studied by measuring plasma parameters with the help of a planar Langmuir probe and emissive probe. The increase in plasma potential, electron temperature and plasma density is understood by loss of bulk plasma electrons to biased disk electrode and the role played by primary electrons in the ionization process. A detailed description of the experimental set-up and plasma production is given in Sec.\ref{sec:exp_setup}. The variation of plasma parameters with bias voltage on the disk electrode and filament heating currents (primary electron population) are discussed in Sec.\ref{sec:exp_results}. The observed results are discussed in Sec.\ref{sec:results_discussion}. A brief summary of the work along with concluding remarks is provided in Section~\ref{sec:summary}.
\section{Experimental Setup} \label{sec:exp_setup}
\begin{figure*} 
\centering
 \includegraphics[scale= 0.8000]{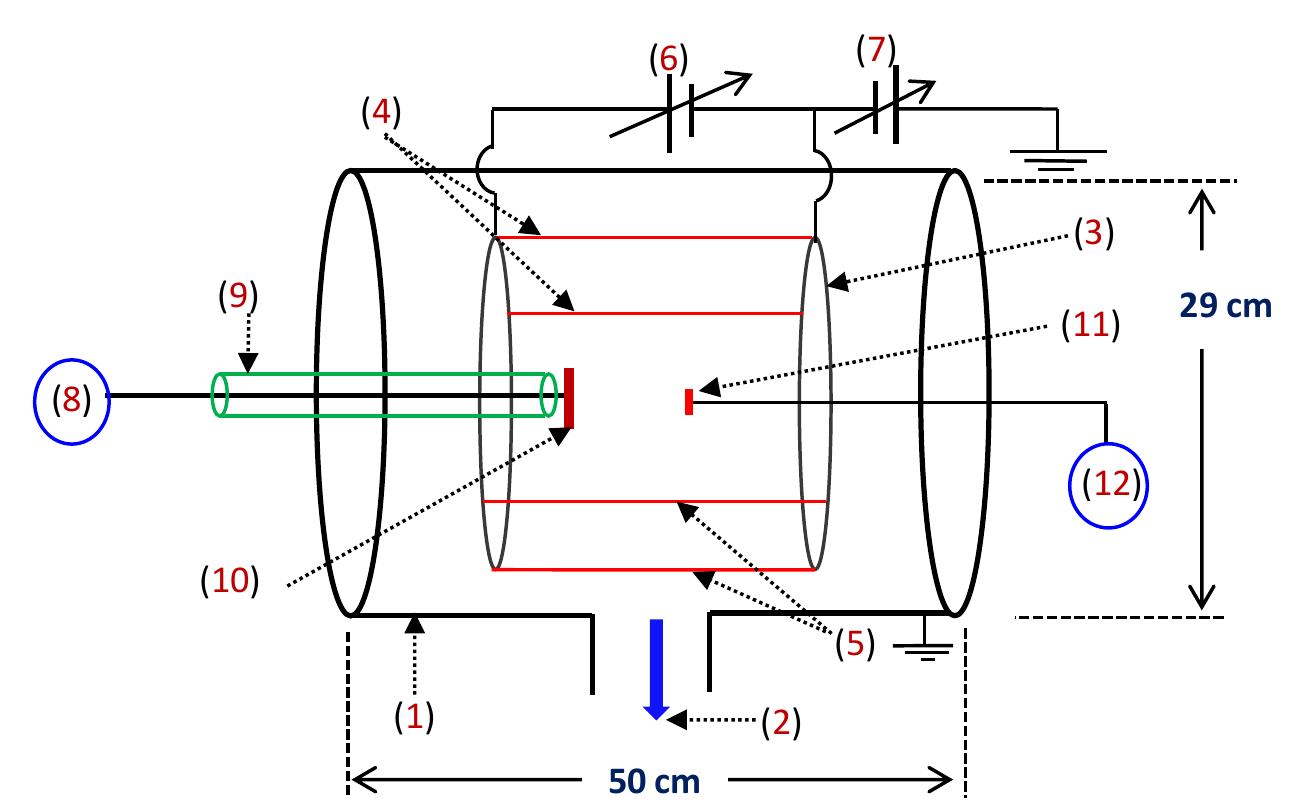}
\caption{\label{fig:fig1}Schematic of the experimental set-up: (1) Experimental vacuum chamber, (2) pumping assembly, (3) stainless steel ring, (4) and (5) thoriated tungsten filaments, (6) filament heating supply, (7) discharge power supply, (8) disk bias supply, (9) ceramic tube, (10)  metallic SS disk, (11) Planar Langmuir probe and (12) Langmuir probe supply.}
\end{figure*}

 The experiments were conducted in a grounded cylindrical chamber of stainless steel (SS) having an inner diameter of 29 cm and length of 50 cm. The vacuum chamber was evacuated with the help of a pumping assembly consisting of a combination of rotary pump and diffusion pump. The chamber was evacuated to base pressure of $\sim$ $1\times 10^{-5}$ mbar. The pressure inside the vacuum chamber was measured by a Pirani gauge. A Needle valve attached to the chamber was used to regulate the gas pressure inside the vacuum chamber. The plasma is generated by electron impact ionization of helium atoms at working pressure of $1\times 10^{-3}$ mbar by primary energetic electrons (60 to 80 eV) emitted from four DC biased hot thoriated tungsten filaments of radius 0.125 mm. These filaments were mounted on two SS rings of 12 cm radius. The filaments were heated by using a direct current power supply (32 V, 30 A) and then biased to a potential of -65 V with respect to the grounded chamber by using a discharge power supply (0--300 V, 5 A). Since electrons emitted from the heated filaments (thermionic electrons) have very low energy (0.2 to 0.3 eV), an external negative biasing is required to accelerate the thermionic emitted electrons above the ionization potential of helium gas ($\sim$ 24.58 eV). A one-sided stainless steel disk (or auxiliary electrode) of diameter 3 cm or 8 cm was used to modify the steady-state low-pressure plasma. A DC voltage supply was used for biasing the metal disk from -50 V to +50 V. A one sided-planar Langmuir probe \cite{probereview1,chenprobe1,mangilalthesis} of radius 4 mm was used to measure the plasma density ($n$) and electron temperature ($T_e$). The plasma potential at a given experimental condition was measured using the emissive probe \cite{emissivesheehan}. Schematic diagram of the experimental setup is given in Fig.\ref{fig:fig1}. \par
\begin{figure}
 \includegraphics[scale= 0.3505000]{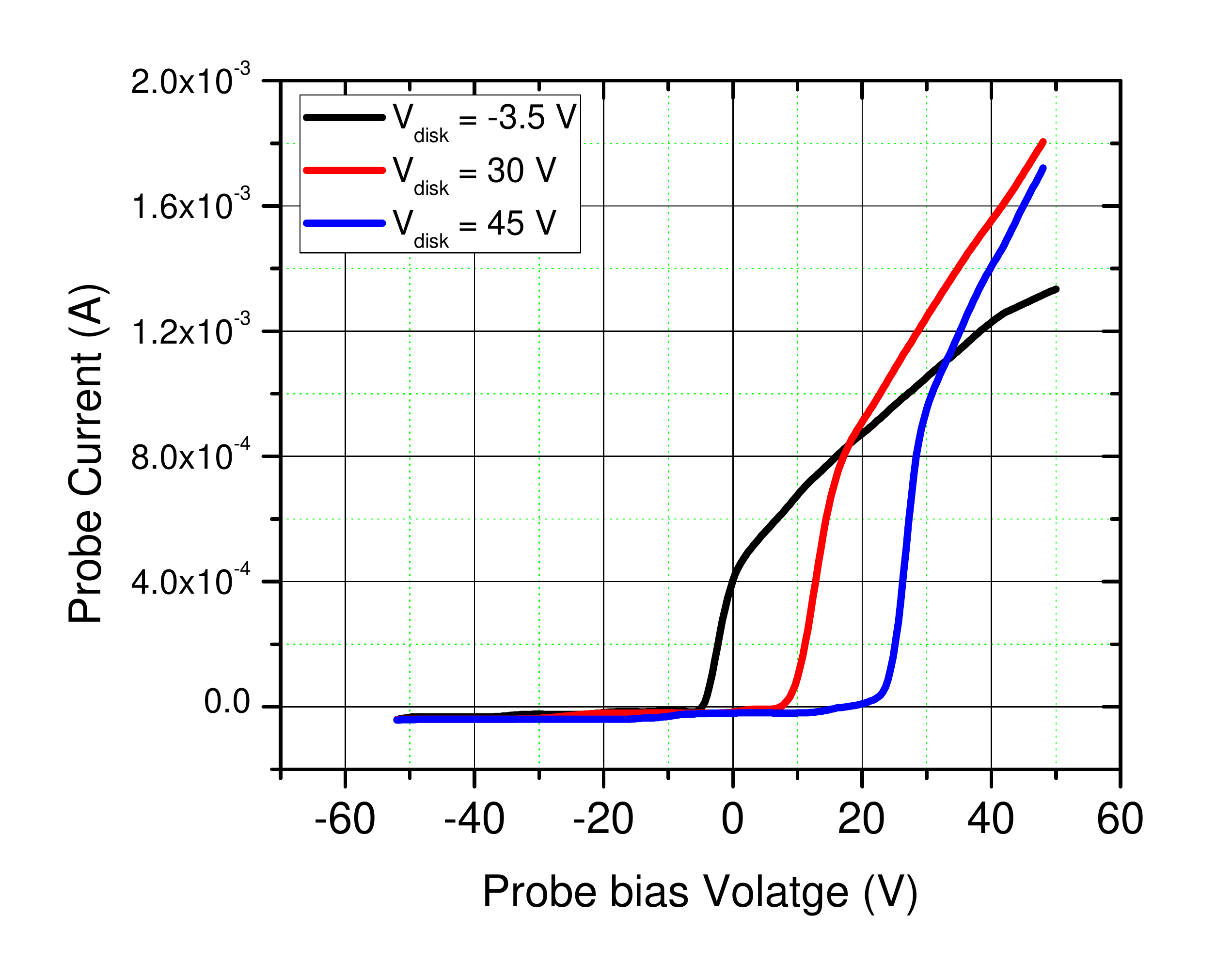}
\caption{\label{fig:fig2}I--V characteristics of the planar probe (smoothed) at three different bias voltages to the disk electrode. The experiment was performed in helium plasma at pressure $1\times 10^{-3}$ mbar and discharge current, 50 mA. Diameter of the disk electrode was 3 cm and probe was kept at 10 cm away from the biased metal disk on the discharge axis.}
\end{figure}
\section{Experimental Results} \label{sec:exp_results}
To see the effect of either positively or negatively biased auxiliary (additional) metal electrode (disk) on the low-pressure filament discharge, plasma parameters ($n_e, T_e$ and $V_p$) were measured using the electrostatic probes. A stainless steel metal disk of diameter, D = 3 cm or 8 cm and thickness 0.5 mm was placed at a fixed position along the discharge axis (see Fig.\ref{fig:fig1}). The backside of the disk was covered by a dielectric (Nylon) material and only the front side was exposed to helium plasma. A planar probe at 10 cm away from the metal disk on the same axis was used to get the current-voltage characteristics (I--V curve) by applying a voltage ramp of range -50 to 50 V. The plasma potential at same position was measured by a radial moving emissive probe. We also measured the plasma potential using the cold probe technique (first derivative of I--V curve of planar probe) for the given experimental conditions. Difference in the measured plasma potential by both hot and cold probe techniques was $<$ 5\%. For getting the aimed data, metal disk was biased in the potential range of -30 V to +50 V and the corresponding probe current was measured to construct the I--V characteristics for a given disk bias voltage. \par
For the first set of experiments, helium gas pressure and discharge current were kept fixed at $1 \times 10^{-3}$ mbar and 0.05 A respectively. The discharge current depends on the filament heating current (or thermionic emitted current flux) and negative DC bias voltage applied to heated filaments. In this case, filaments were heated by passing the current of 24 A, and thermionic emitted electrons were accelerated by applying the negative bias of 65 V to filaments. The I--V characteristics of the planar probe while the metal disk (D = 3 cm) was kept at -3.5 V, 30 V, and 45 V are depicted in Fig.\ref{fig:fig2}. These I--V characteristics at fixed discharge conditions indicate change in the plasma characteristics after immersing or placing a positively biased auxiliary metal disk in the plasma column. For getting the variation of plasma parameters against metal disk biasing at a given discharge condition, the I--V curves taken at various disk bias voltages were analyzed.\\
\begin{figure*} 
 \centering
\subfloat{{\includegraphics[scale=0.36050]{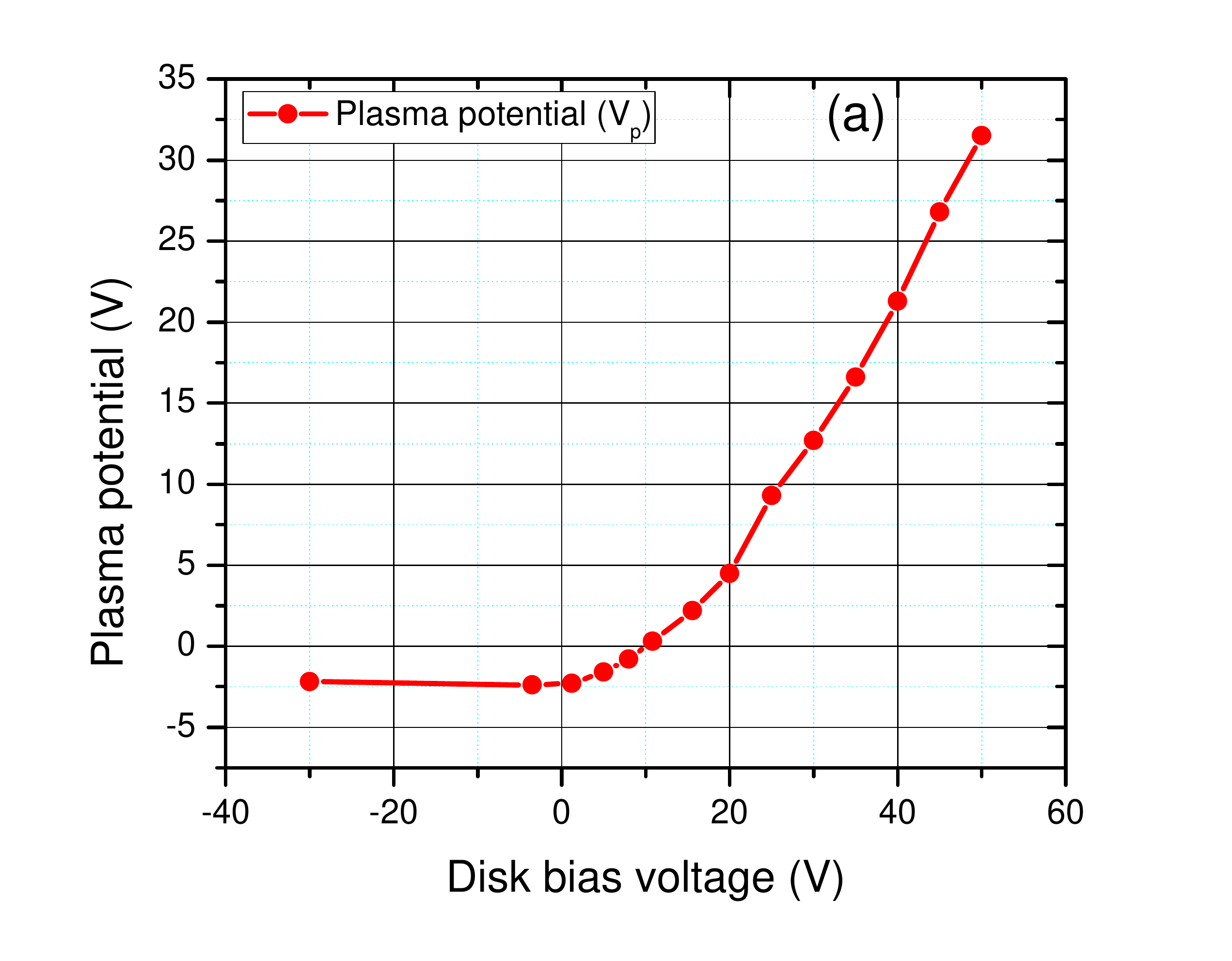}}}%
\hspace*{-0.5in}
 \subfloat{{\includegraphics[scale=0.36050]{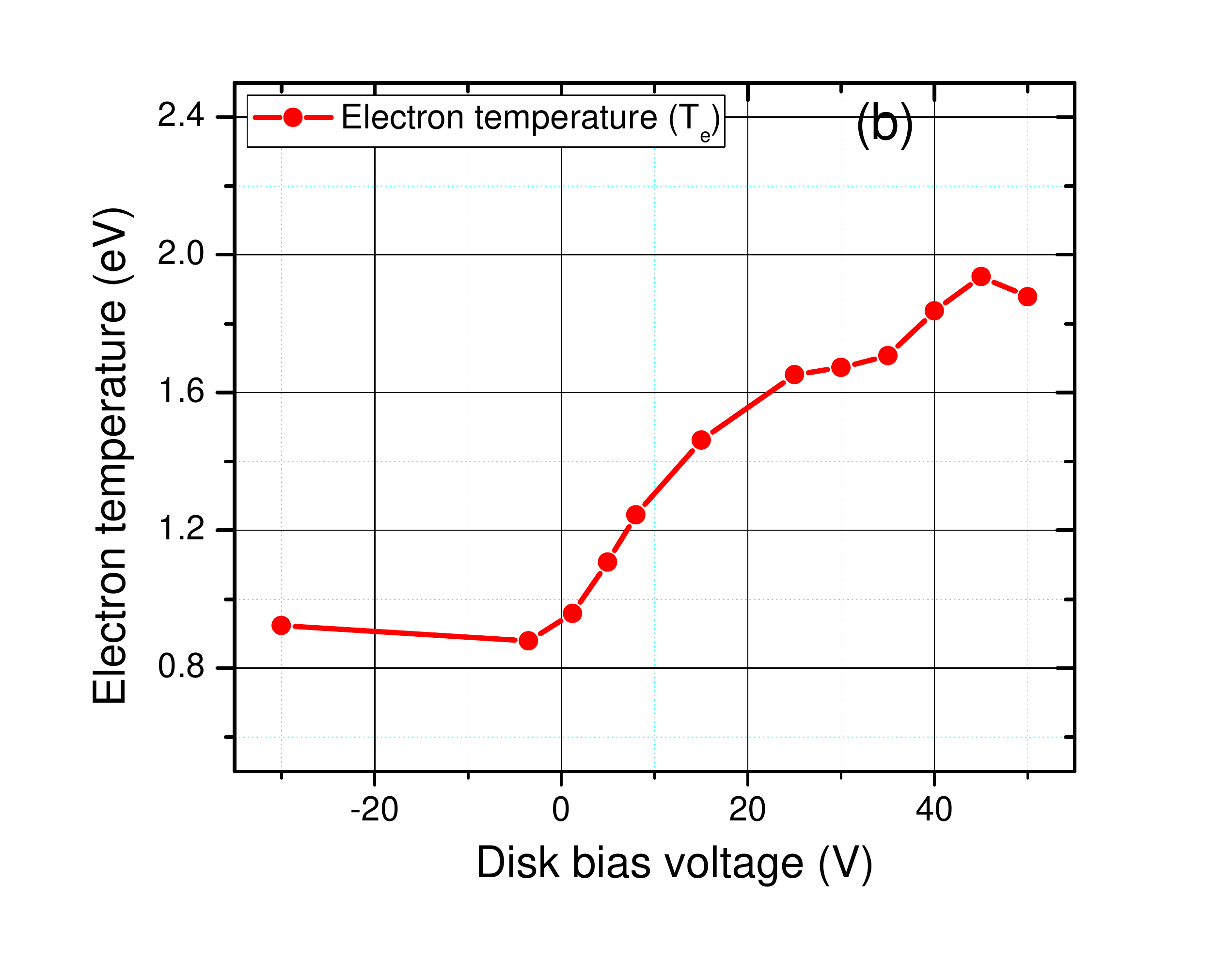}}}
 \qquad
 \subfloat{{\includegraphics[scale=0.36050]{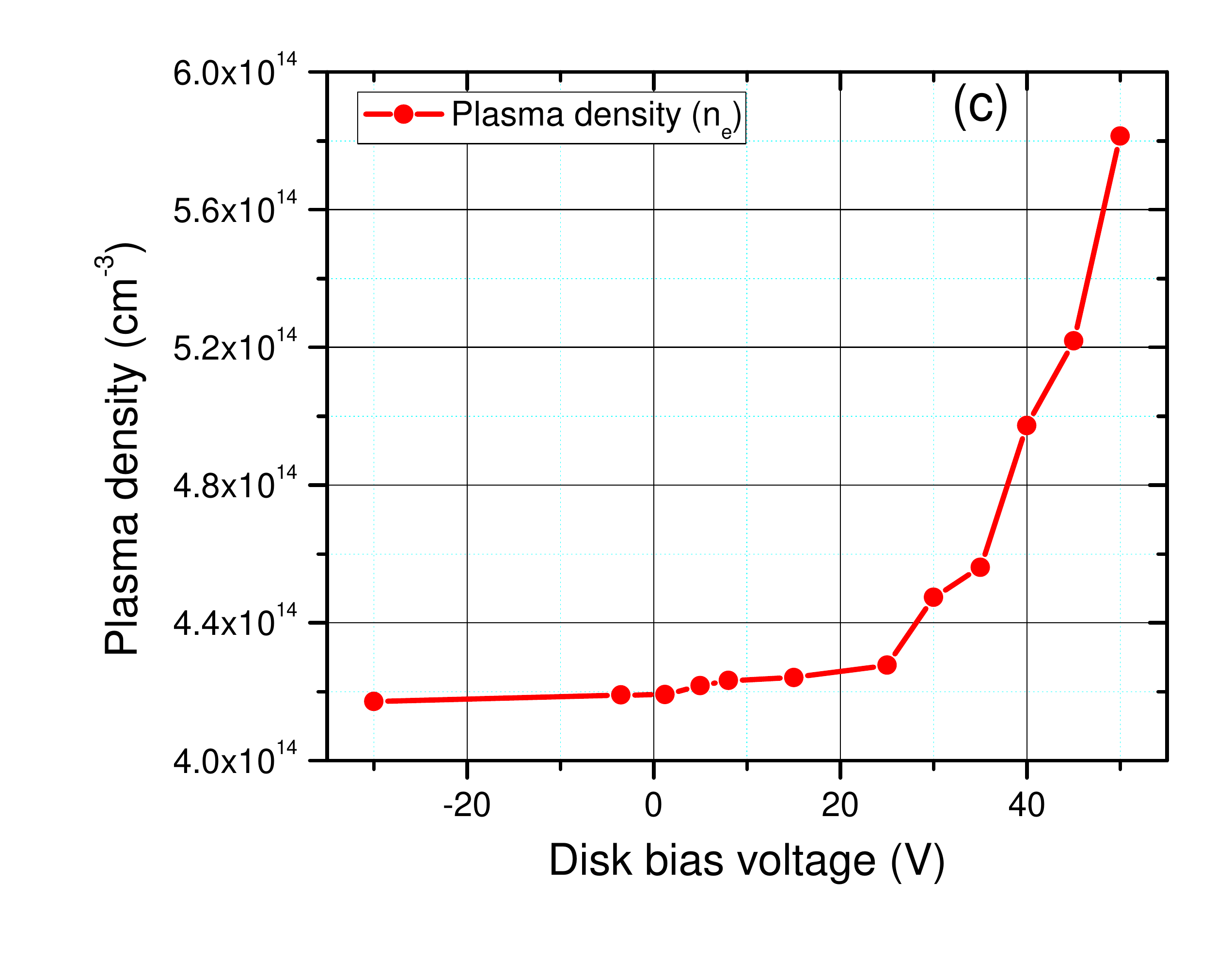}}}
\caption{\label{fig:fig3}(a)Plasma potential variation, (b) Electron temperature variation and (c) plasma density variation against metal disk bias voltage. The helium pressure was set at $1\times 10^{-3}$ mbar and filament heating current was 24 A. The diameter (D) of the metal disk was 3 cm. Error over the averaged value of the measure plasma parameters are $< \pm$ 7\%.} 
\end{figure*}
\begin{figure*}
\centering
 \includegraphics[scale= 0.36000]{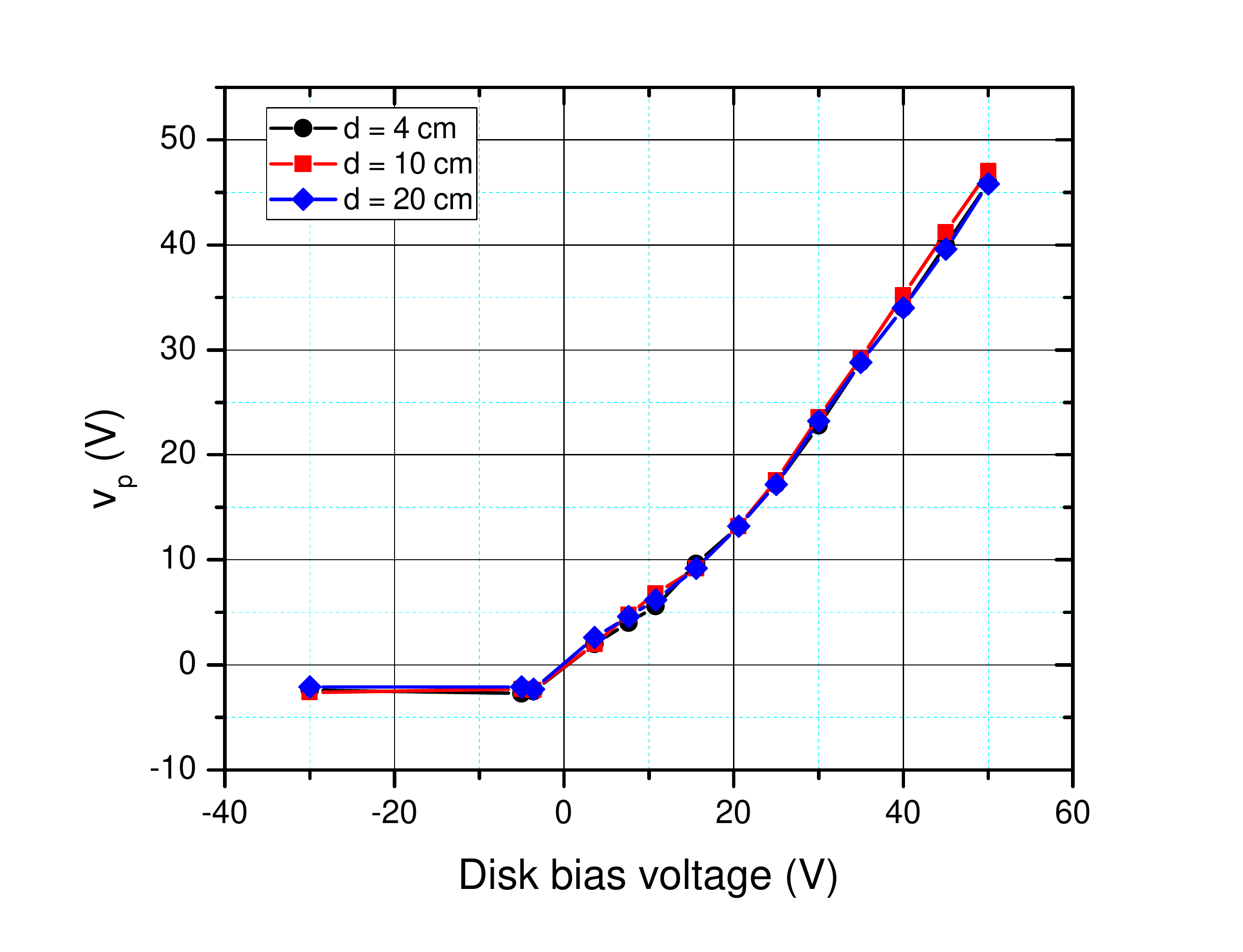}
\caption{\label{fig:fig4}The plasma potential ($V_p$) variation with the disk bias voltage at three different locations from the biased disk. The helium gas pressure was set at $1\times 10^{-3}$ mbar and filament heating current was 24 A. The diameter (D) of the metal disk was 8 cm. Error over the averaged value of the measure plasma potential is $< \pm$ 5\%. }
\end{figure*}
\begin{figure*} 
 \centering
\subfloat{{\includegraphics[scale=0.3550]{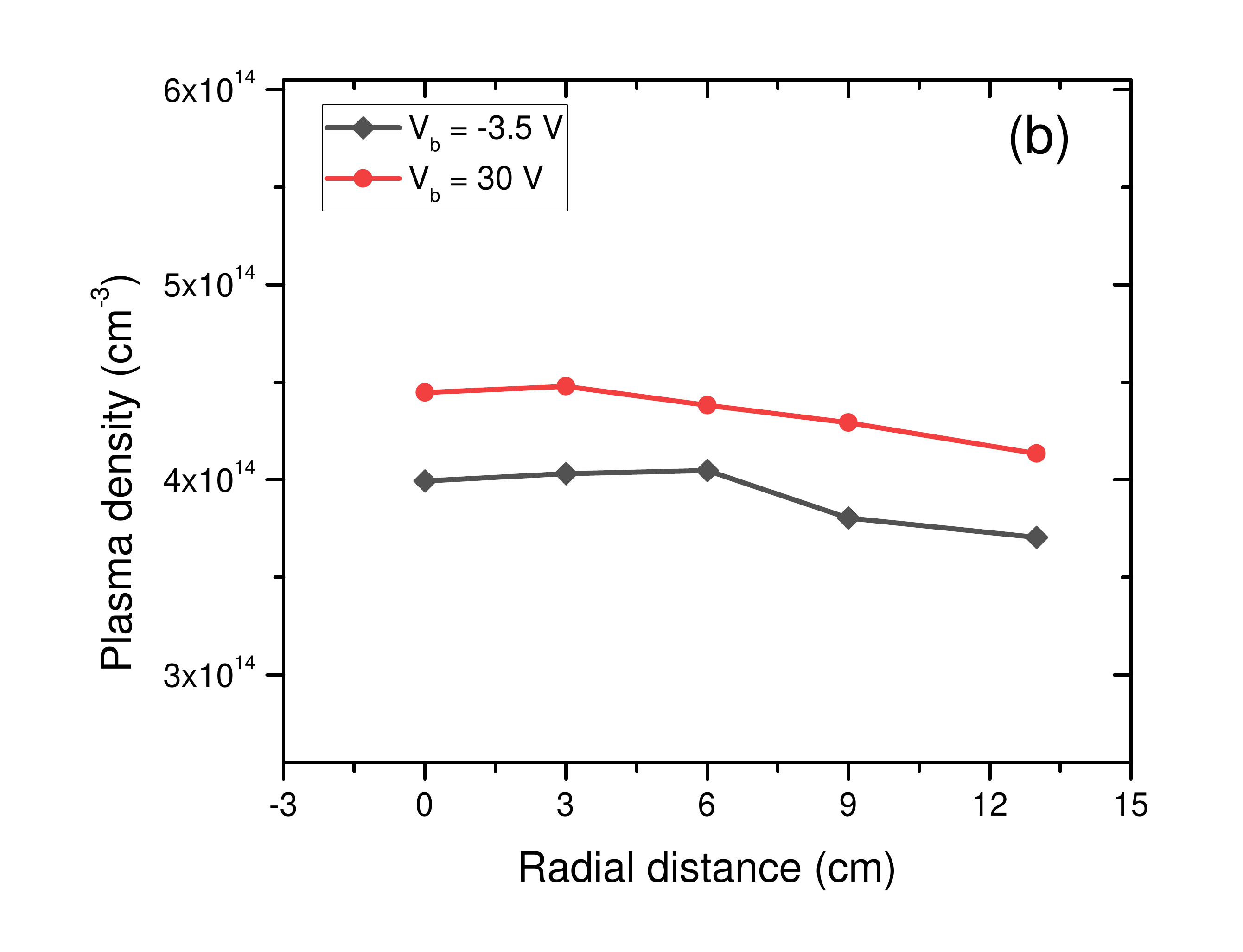}}}%
\hspace*{-0.5in}
 \subfloat{{\includegraphics[scale=0.3550]{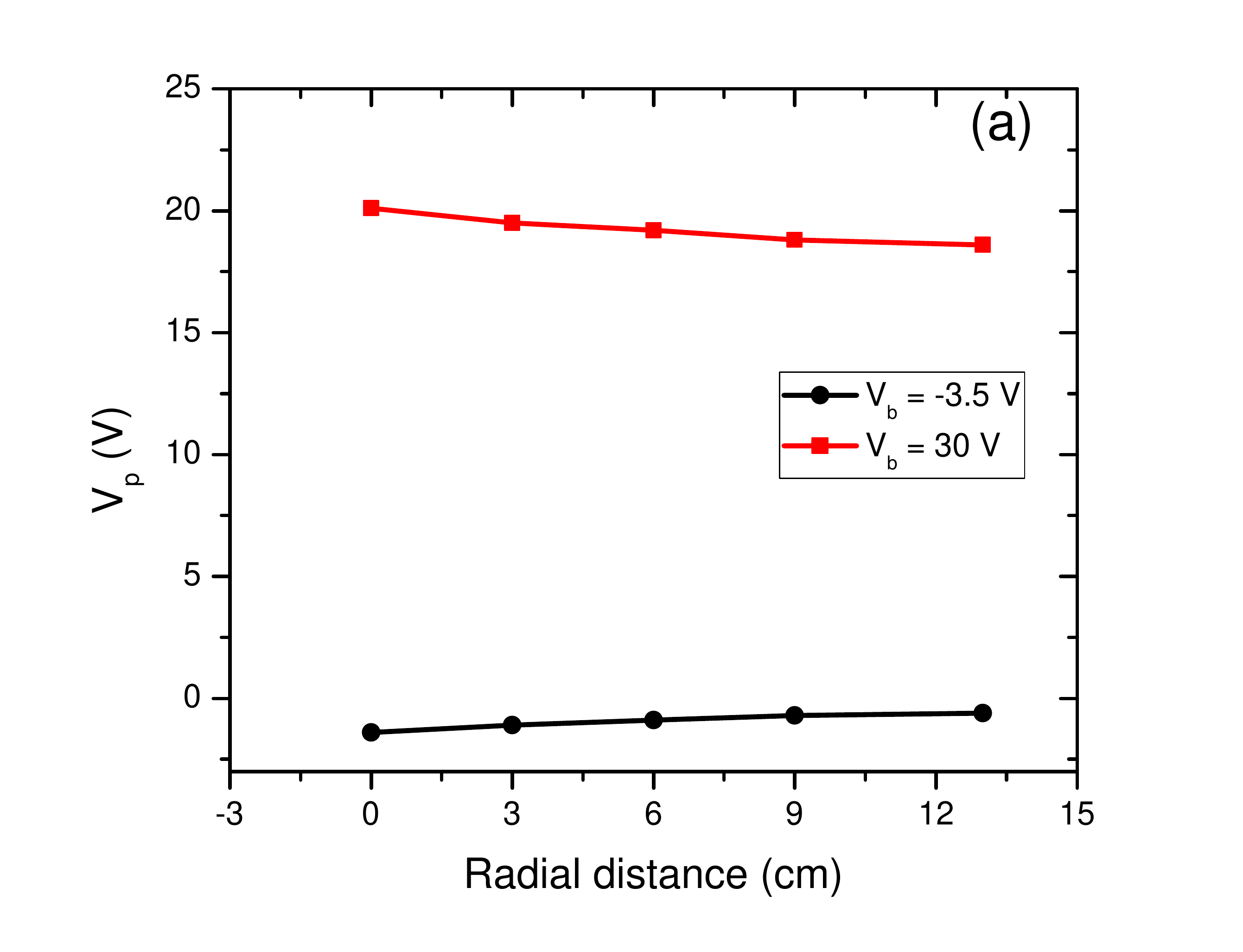}}}
 \qquad
\caption{\label{fig:fig5}(a) The radial plasma potential and (b) plasma density variation against the disk bias voltage. The measurements were taken at d = 10 cm. The helium gas pressure was set at $1\times 10^{-3}$ mbar and filament heating current was 24 A. The diameter (D) of the metal disk was 8 cm. Error over the averaged value of the measure plasma potential is $< \pm$ 5\%.} 
\end{figure*}
In Fig.\ref{fig:fig3}(a), we have plotted the plasma potential variation against the disk bias voltages. The plasma potential varies slowly at low positive bias voltage ($<$ 20 V) and has an approximate linear growth above +20 V. The measured plasma potential is always found to be lower than the disk bias voltage. The negative value of plasma potential is expected due to the presence of primary energetic electrons (beam electrons) in the plasma volume. At low disk bias voltage, it is difficult to determine the population of these beam electrons but can be possible at large positive bias voltage on the disk electrode \cite{mangilalcpp}. The increase in plasma potential (towards more positive) is expected by removing the confined electrons of the plasma medium \cite{particlebalancemodel1,particlebalancemodel3exp,anodesizeplasmareponse}. 
The rate of change of $V_p$ depends on the background plasma density, population of energetic electrons, and size of the auxiliary disk electrode. Similar to the plasma potential, the plasma density ($n_e$) and electron temperature ($T_e$) were calculated using the recorded I--V data at various metal disk bias voltages. The $T_e$ variation with the disk bias voltages is given in Fig.\ref{fig:fig3}(b). We observe an increment in $T_e$ with increasing the positive bias voltage on the disk which is expected in case of higher positive potential \cite{particlebalancemodel1,particlebalancemodel2,particlebalancemodel3exp,
anodesizeplasmareponse}. The plasma density variation against disk bias voltages is shown in Fig.\ref{fig:fig3}(c). In Fig.\ref{fig:fig3}(c), we can see that $n_e$ increases with increase in the disk bias voltage above the plasma potential. At large positive bias voltage (+50 V), plasma density is observed to be maximum. Thus, we observe the plasma density enhancement from its equilibrium (initial) value in the presence of steady state perturbation applied by a positively biased metal disk electrode. \par
In the next set of experiments, we carried out the experiments with a large diameter disk electrode (D = 8 cm) at same discharge conditions. The plasma potential with disk bias voltages at three different locations from the biased disk (d = 0 cm). For this biased disk electrode (D = 8 cm), the ratio $A_d/A_w < \mu$ which suggests a slow variation of plasma potential with biasing voltage to the disk \cite{globalambipolardiffusion,plasmapotentiallocking}. However, the plasma potential increases at a higher rate than that expected from the theoretical model as we increase the bias voltage to metal disk (see Fig.\ref{fig:fig4}). The plasma potential profiles are nearly same at different locations from biased disk which confirms the effect of the positively biased electrode on the entire bulk plasma (or plasma column). The plasma density and $T_e$ increase with increasing the bias voltage to disk electrode and found a nearly similar trend that of observed for small-sized biased metal disk (see Fig.\ref{fig:fig3}). With the same experimental configuration, we also measured plasma potential and plasma density radially from the discharge axis ($r$ = 0 cm) at d = 10 cm. In fig.\ref{fig:fig5}, we see a slight change in $V_p$ and $n$ radially. It shows that the effect of biased disk electrode is seen throughout the bulk plasma.\par
\begin{figure*}
 \centering
\subfloat{{\includegraphics[scale=0.3550]{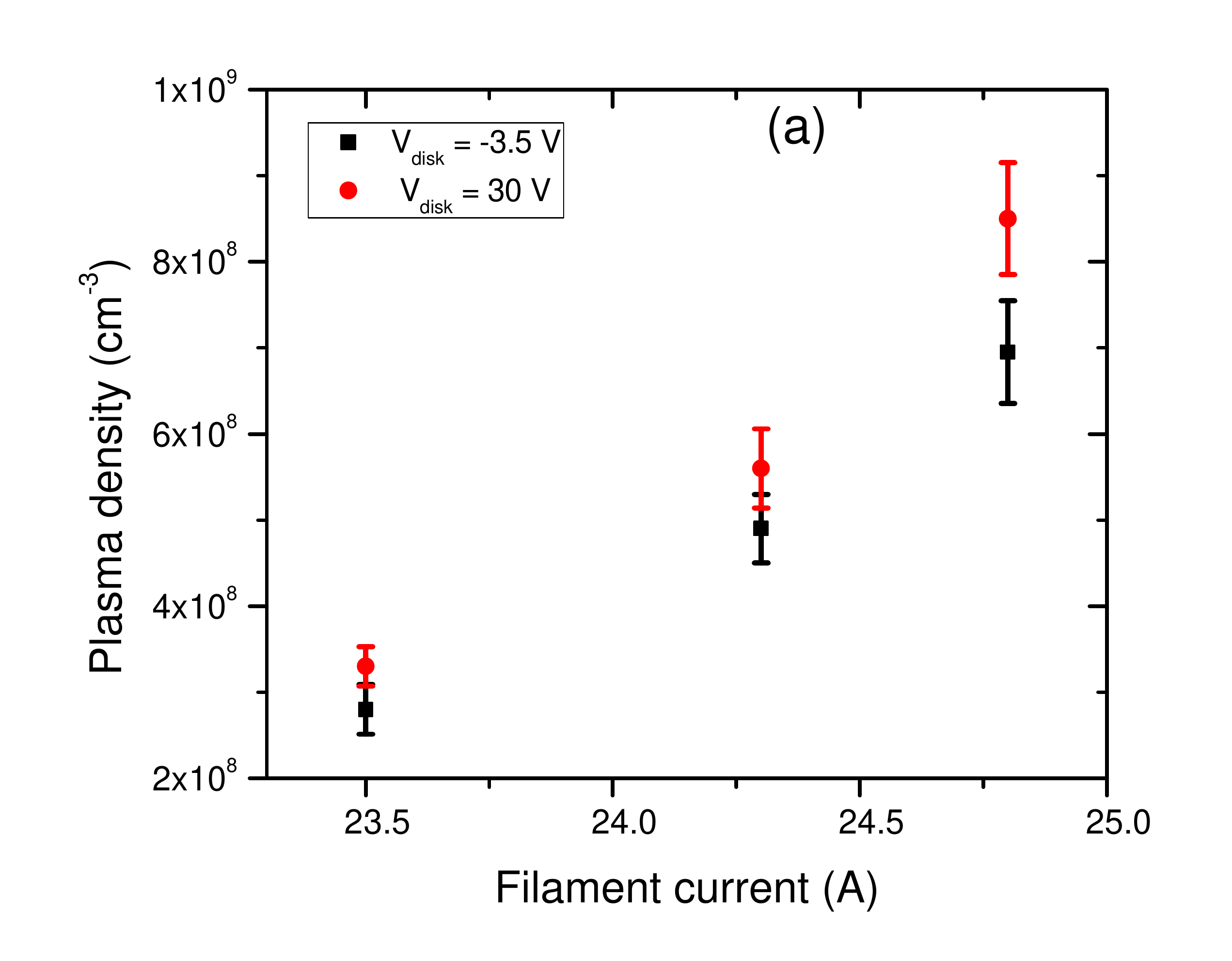}}}%
\hspace*{-0.3in}
 \subfloat{{\includegraphics[scale=0.3550]{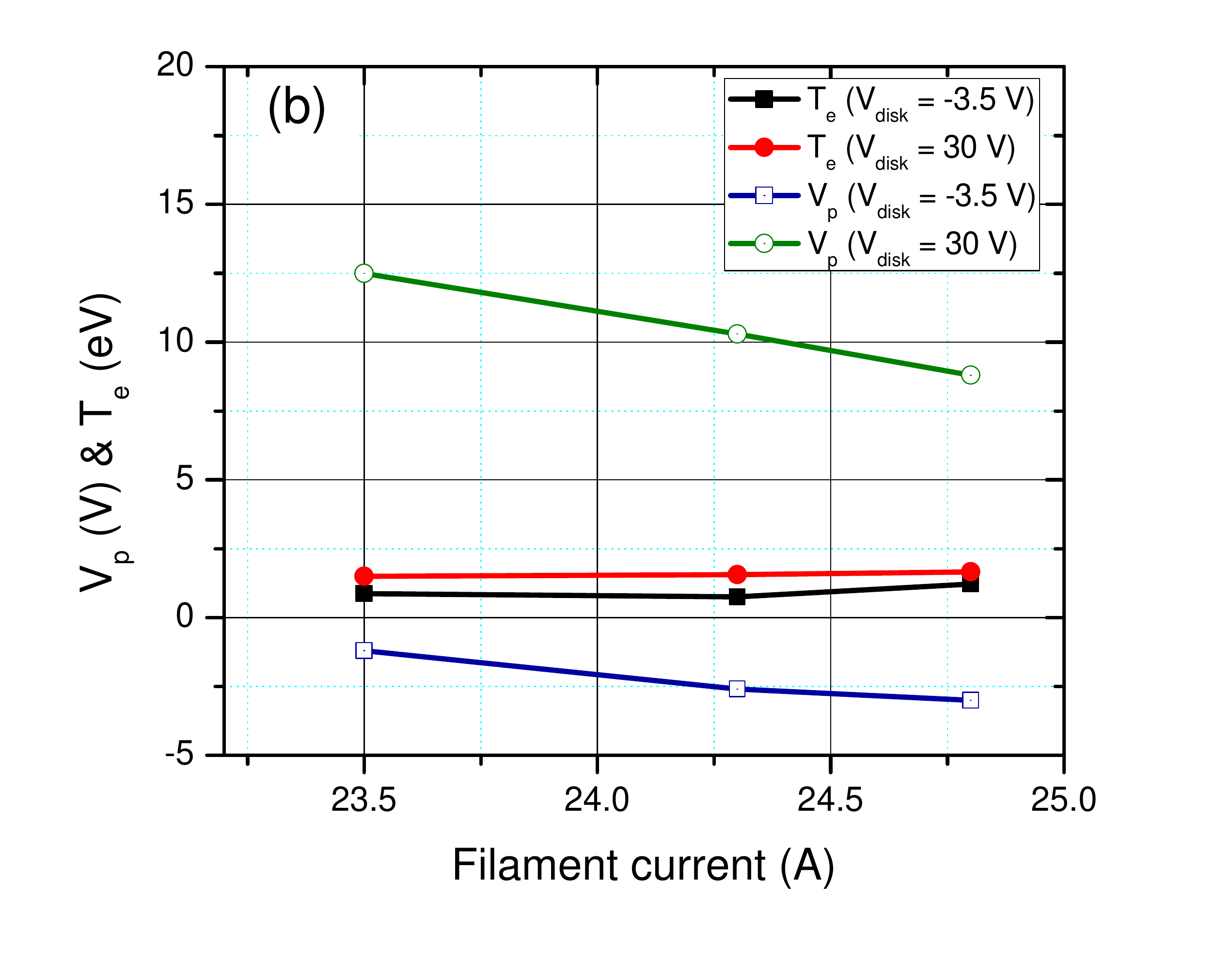}}}
\caption{\label{fig:fig6}(a) Plasma density variation with filament heating current at two disk bias voltages. (b) Electron temperature and plasma potential variation with filament heating current at two disk bias voltages. The helium pressure was set at $1\times 10^{-3}$ mbar. The diameter (D) of the metal disk was 3 cm. Error over the averaged value of the measure plasma parameters are $< \pm$ 5\%.} 
\end{figure*}

 
The last set of experiments was conducted to see the effect of primary energetic electron flux on the plasma parameters when an additional positively biased metal disk is immersed in the plasma column (see Fig.\ref{fig:fig1}). The I--V characteristics of probe were recorded for two disk bias voltages (-3.5 and +30 V) at different filament currents (or thermionic electron currents). The primary energetic electrons flux increases with increasing the filament current at a fixed DC bias voltage (-65 V). Variation of plasma density with filament heating currents for two disk bias voltages is shown in Fig.\ref{fig:fig6}(a). The plasma density increases if flux of primary energetic electrons is increased by passing more current through the filaments. The plasma density increment rate depends on the bias voltage of the metal disk, as can be seen in Fig.\ref{fig:fig6}(a). Apart from density variation plots, we have also plotted $V_p$ and $T_e$ variation against the filament currents in Fig.\ref{fig:fig6}(b). It is clear from fig.\ref{fig:fig6}(b) that $T_e$ has different values for bias voltage -3.5 V and +30 V for a given filament current but a slight increment in $T_e$ was observed with increasing the population of primary energetic electrons or filaments heating. We observe slight lower plasma potential at $V_b$ = 30 V as filament heating current is increased (see Fig.\ref{fig:fig6}(b)). It shows that plasma potential is going to decrease with increase in the population of primary energetic electrons. We expect similar behavior of $V_p$ against filament heating current if the metal disk is positively biased above the plasma potential. This behavior suggests the role of primary energetic electrons (or electrons beam) on plasma characteristics in the presence of an auxiliary positively biased metal electrode (disk). A detailed discussion on the observed experimental results is provided in Sec.\ref{sec:results_discussion}.  
\section{Discussion}  \label{sec:results_discussion}
In a steady-state low-pressure filament discharge, the plasma production rate is balanced by the loss rate to the chamber wall. The plasma production rate depends on the energetic emitted electron flux from the negatively biased heated filaments\cite{particlebalancemodel1,particlebalancemodel2,particlebalancemodel3exp}. It was observed that the plasma characteristics get modified in the presence of a positively biased auxiliary metal electrode (disk) in the plasma column. For a negatively biased disk electrode in plasma, an ion sheath is formed in front of it which reduces the electron flux and increases the ion flux towards the disk. Since this field is localizing up to a few mm distance and after that its effects are not dominating inside the plasma, the plasma parameters are not changing with the negatively biased disk at a given discharge condition. If we give positive bias to the metal disk then an electron or ion sheath is formed around the positively biased metal disk in the hot-filament discharge \cite{globalambipolardiffusion,plasmapotentiallocking}. In such a case, positively biased metal electrode perturbs the entire plasma and influence the bulk plasma properties. As disk bias voltage is kept above the initial plasma potential, the confined electrons (plasma electrons) are lost to the surface of the disk. Due to the loss of plasma electrons, plasma potential becomes more positive to maintain the quasi-neutrality condition of plasma. However, the plasma potential variation of bulk plasma with the disk bias voltage depends on the area of the biased metal disk ($A_d$) in order to maintain current balance \cite{globalambipolardiffusion,plasmapotentiallocking}. The response of the bulk plasma to a biased metal electrode above the plasma potential is mainly determined by the ratio of area $A_d/A_w$ and the ratio of ion-electron mass $(M_i/m_e)$.  The ratio of area $A_d/A_w$ determines whether an electron sheath ($A_d/A_w$ $< \mu$) or a double sheath ($A_d/A_w$ $\simeq$ 1.7 $\mu$) or ion sheath ($A_d/A_w$ $>$ 1.7 $\mu$) will be formed around the positively biased ($V_b > V_p$) auxiliary electrode (disk) in the helium plasma \cite{plasmapotentiallocking,globalambipolardiffusion,anodesizeplasmareponse}. Here $\mu$ = $\sqrt{2.3 m_e/M_i}$ and $V_b$ is the disk bias voltage. For helium gas, $\mu$ is $\sim$ 1.76 $\times 10^{-2}$. In the present experiment, the area of the plasma-facing chamber wall $A_w$ $\sim$ 6350 $cm^2$ and area of disk electrode, D = 3 cm and D = 8 cm  are $A_d$ $\sim$ 7 $cm^2$ and $A_d$ $\sim$ 50 $cm^2$ respectively. So the ratio $A_d/A_w$ $< \mu$ for both biased disks which predicts the electron sheath near the positively biased disk \cite{globalambipolardiffusion}. Since the characteristics of bulk plasma depends on the type of sheath region around a positively biased electrode, plasma parameters ($n_e$, $T_e$ and $V_p$) should also vary according to that of predicted in the theoretical model as well as in experiments \cite{globalambipolardiffusion,anodesizeplasmareponse,plasmapotentiallocking}.\par
The plasma potential variation with the bias voltages to the metal disk, Fig.\ref{fig:fig3}(a) and Fig.\ref{fig:fig4}, does not follow the trend (sub-linear) that were predicted by theoretical model if electron sheath is formed ($A_d/A_w$ $< \mu$) in front of metal disk. The plasma potential increases with a higher rate (super-linear) when we give more positive bias voltage ($>$ +20 V) to the metal disk. Such super-linear variation of plasma potential is expected for a large sized biased disk electrode ($A_d/A_w$ $\geq$ 1.7 $\mu$).
The electron temperature of bulk plasma ($T_e$) also increases in the presence of a positively biased auxiliary metal disk. The variation of $T_e$ is somewhat similar to that predicted in theory and experiment \cite{globalambipolardiffusion,anodesizeplasmareponse}. There is a major difference in the plasma density variation with bias voltages above the plasma potential in the case of $A_d/A_w$ $< \mu$. In our experiments, we have observed the enhancement of plasma density (see Fig.\ref{fig:fig3}(c)) instead of electron density depletion \cite{anodesizeplasmareponse,globalambipolardiffusion} in the presence of a smaller sized positively biased auxiliary metal electrode (disk). The plasma density increases linearly above the disk bias voltage of 25 V. This opposite behavior of plasma density variation could be due to the presence of primary energetic electrons (electron beam) in the bulk plasma.\\
Plasma response to a fixed biased disk (+30 V) at different fluxes of primary energetic electrons (heating currents) may help in understanding the observed results. We see an effective change in plasma density (see Fig.\ref{fig:fig6}(a)) for the negatively biased (-3.5 V) and positively biased (+30 V) disk at a large filament heating current (24.7 A). It indirectly shows the role of the population of primary energetic electrons (or thermionic electrons) on the plasma properties in the presence of a positively biased disk. In Fig.\ref{fig:fig6}(b), the gap between plasma potential measured at $V_b$ = -3.5 V and +30 V is decreased by $\sim$ 2 V if filament current is increased by 1.2 A (see Fig.\ref{fig:fig6}(b)). It indicates that the beam electrons (primary energetic electrons) lower the plasma potential that was also observed in other hot-filament discharge\cite{negativeplasmapotentialcpp}. \par
The observed results can also be understood qualitatively based on the past theoretical and experimental work in the low-pressure hot-filament discharge \cite{plasmapotentiallocking,particlebalancemodel1,particlebalancemodel2,particlebalancemodel3exp,mangilalcpp}. Once the disk bias voltage is higher than the plasma potential ($V_b > V_p$), bulk plasma electrons loss rate to the metal disk increases, resulting in increasing the plasma potential. With increasing the plasma potential by $\sim$ 30 V (Fig.\ref{fig:fig3}(a)), the energy of thermionic emitted electrons is increased by $\sim$ 30 eV. These energetic primary electrons from heated filaments (90 to 100 eV) create the plasma by the impact ionization of helium gas. Since the chamber wall is grounded ($V_w$ = 0), the bulk plasma electrons are confined in a potential well which is determined by the plasma potential. As the disk bias voltage is above 25 V, plasma density increases at a higher rate. It is only possible with a higher ionization rate in the bulk plasma volume. According to the theoretical model\cite{particlebalancemodel1,particlebalancemodel2,particlebalancemodel3exp,secondaryelectronsinplasma}, secondary electrons (hot confined electrons) which are released from the chamber wall and produced from the primary energetic electrons are also confined in the potential well. The secondary electrons of kinetic energy (E $>$ 25 eV) above the ionization potential of helium gas ($\sim$ 24.5 eV) may also take part in the ionization process and increase the plasma density. If we increase the density of the primary energetic electrons, density growth rates are different for negatively (-3.5 V) and positively (+30 V) biased disk (see Fig.\ref{fig:fig6}(a)). It confirms that the density of primary energetic electrons of the same energy plays a dominant role in determining the plasma formation rate (or ionization rate) in presence of a positively biased auxiliary metal disk.\par
The positively biased disk ($V_b$ $>$ $V_p$) attracts the low-energy plasma electrons therefore average energy of the bulk plasma electrons is increased\cite{anodesizeplasmareponse,globalambipolardiffusion}. It is also expected to increase the average energy bulk plasma electrons by collisions of low energy electrons and secondary high energy electrons in the potential well \cite{robertsonincreasete,particlebalancemodel3exp,secondaryelectronsinplasma}.
Hence we observe increment in electron temperature ($T_e$) of the bulk plasma with increasing the disk bias voltage.
\section{Summary}  \label{sec:summary}
In this work, we have seen the response of plasma with an auxiliary (additional) biased disk electrode immersed in the hot-filament low-pressure helium plasma. The plasma parameters ($V_p$, $T_e$ and $n$) were measured using the planar and emissive probe at various locations of the plasma column. For both sizes of the metal disk (D = 3 cm and D = 8 cm), plasma potential, plasma density, and electron temperature were observed to increase with increasing the bias voltage to the metal disk above the plasma potential at given discharge conditions. However, the rate of change of plasma parameters with disk bias voltage depends on the size of the disk electrode and equilibrium plasma density (or population of primary energetic electrons). Such response of helium plasma is understood with the help of available theoretical models. The ionization rate (plasma production rate) and plasma loss rate to available boundaries mainly determine the characteristics of bulk plasma. Discrepancy in the observed results (plasma parameters variation) from that of predicated or observed in hot-filament discharge could be possible due to the presence of energetic primary electrons in the bulk plasma. The experimental results suggest to incorporate the role of primary energetic electrons emitted from heated filaments in the available theoretical model. These experimental results also provides an insight into electrostatic confinement by using an positively biased additional electrode. In future, our focus will be to study the transient response of hot-filament plasma at different electrostatics confinement strengths which depends on the height of potential well. The height of potential well can be modified by changing the plasma potential with respect to grounded chamber.\\ 
\section{Acknowledgement} 
The authors are very grateful to Prof. S. Mukherjee for guiding during the experiments at Institute for Plasma Research, Gandhinagar.
\bibliography{biblography}
\end{document}